\documentstyle[prbbib,preprint,aps]{revtex}
\begin{document}
\draft
\title{Theory of adsorption and desorption of H$_2$/Si(001)}
\author{E. Pehlke and M. Scheffler}
\address{Fritz-Haber-Institut der Max-Planck-Gesellschaft\\
Faradayweg 4-6, D-14195 Berlin (Dahlem), Germany}
\maketitle

\begin{abstract}

While the small sticking coefficient for molecular
hydrogen on the Si(001) surface apparently requires
a large energy barrier of adsorption, no such
barrier is observed in desorption experiments.
We have calculated the potential-energy surface of an H$_2$
molecule in front of a Si(001) surface.
If we relax the Si substrate, we find an optimum desorption path
with a low ($\lesssim$~0.3 eV) adsorption energy barrier.
While molecules impinging on the surface will mostly
be reflected at the larger barrier of some frozen-substrate,
molecules adsorbed on the surface can desorb along
the low-barrier path.

\end{abstract}


Adsorption and desorption processes
represent the initial and final step of catalytic reactions
of gases on solid surfaces, which are of importance
both in fundamental research and for technological applications.
Typically, the adsorption and desorption dynamics
of diatomic molecules are
described within a simple conception that the molecule is
moving in a fixed, elbow shaped potential $E(Z,d)$, with $Z$
denoting the distance of the molecule from the surface, and $d$
being the separation of the atoms forming the molecule.\cite{Holloway,book}
Of course this potential may additionally depend on the remaining four
molecular coordinates (i.e., the position
of the center of mass of the molecule along the
surface, $(X,Y)$, and the azimuthal and polar
orientation of the molecular axis),
however this does not alter the basic assumption behind
this potential namely that there are {\em no surface atomic
degrees of freedom}, i.e., the surface atomic geometry is frozen.
Therefore, the trajectories for adsorption and desorption are connected by
time reversal, and, consequently, the height of the adsorption energy barrier
measured in either an adsorption or a desorption experiment has to
be the same.

However, recent experimental results for H$_2$/Si(001)
have revealed a puzzling apparent contradiction to the principle
of microscopic reversibility:\cite{KolasinskiPRL,KolasinskiJCP}
On the one hand, the small sticking coefficient of
molecular hydrogen on Si(001) requires that there is a
substantial energy barrier of dissociative adsorption. On the other
hand, no such barrier is found in associative desorption experiments:
Kolasinski {\it et al.}\cite{KolasinskiPRL} measured the
distribution of the translational, vibrational, and rotational
energy of molecules desorbing from the mono-hydride
surface and found that the hydrogen molecules do not have
any significant energy in access of the thermal energy
corresponding to the surface temperature. From this they
estimated the height of the adsorption energy barrier to be $77\pm80$ meV.

This experimental result severely challenges the widely employed conception
of adsorption-desorption dynamics outlined above.
If the energy barrier were in fact high, some mechanism must be at
work nevertheless allowing the molecules to desorb with a
thermal energy distribution. Extrinsic effects, e.g. the
diffusion of hydrogen atoms to special defect sites
on the Si(001) surface with a low local energy barrier
towards desorption, seem to be at variance with
experimental evidence\cite{KolasinskiJCP}.
Furthermore, no anomalous isotope effect was found\cite{Sinniah}
that would corroborate the model of hydrogen
molecules tunneling through the barrier.\cite{Brenig}
Recently Brenig {\it et al.}\cite{Gross} suggested another
mechanism. Their model potential has an order of 1 eV
barrier, but additionally it possesses a surface oscillator
degree of freedom modelling Si lattice vibrations.
After having crossed the barrier, the desorbing molecule
gets decoupled from the surface, and the excess energy
due to the potential drop behind the barrier cannot
be transferred to the molecule, but Si phonons are excited
instead. Thus, despite the large barrier,
the energy distribution of the desorbing
molecules looks approximately thermal.

Other suggested scenarios start from the assumption that there is no,
or only a small, adsorption barrier in the adiabatic
potential energy surface, in
accordance with the desorption data. Still the
sticking coefficient can be small, if the adsorption
and desorption processes follow different pathways on the potential energy
hypersurface.\cite{KolasinskiPRL,KolasinskiJCP}

Kolasinski {\it et al.}\cite{KolasinskiJCP} have carried out
detailed measurements of the sticking coefficient
of D$_2$ on Si(001) using
molecular beam techniques. They found an increase of the
sticking probability both with nozzle temperature (i.e., with
the energy of the impinging molecules) and with
surface temperature. The increased sticking coefficient
for fast molecules demonstrates that the dissociative adsorption
of hydrogen on Si(001) is activated, and the fit to
the experimental data indicates an average barrier height
of 1 eV, and a large width of the barrier height distribution of
0.6 eV. The fact that sticking is facilitated by surface
temperature corroborates the idea that
surface atom motion is important:\cite{KolasinskiJCP}
There may be certain surface
atom configurations which correspond to low barrier
adsorption pathways, however, heating the surface
is necessary to excite surface atom vibrations that
dynamically generate these configurations.
Cluster calculations have lead to adsorption
energy barrier heights larger than 1 eV,\cite{Jing,Wu}
which appear to be at variance with above explanation.

The purpose of this letter is to
sort out the correct explanation for
the apparent contradiction between the measured adsorption and
desorption dynamics.
We have carried out {\it ab initio} total-energy
calculations\cite{Stumpf} to map the potential energy hypersurface
for a hydrogen molecule in front of a Si(001) surface.
To represent the buckled surface (see Fig. 1) we take a
(2$\times$2) surface unit cell. This is necessary
in order to correctly describe the
ground state of the Si(001) surface with buckled dimers (Fig. 1).
Moreover, we use this unit cell instead of the
smaller (1$\times$2) cell in order to reduce the interaction between
hydrogen molecules in neighboring supercells
through electrostatic and electronic H--H coupling and through
the mechanical relaxation of the substrate.
The Si slab consists of five layers of atoms, the
topmost three of them being relaxed, and the atoms in the
remaining two layers are fixed at their bulk positions.
The dangling bonds on the bottom surface of the slab
are saturated with hydrogen atoms.
The total energy is computed within density functional
theory together with the local density approximation (LDA)
for the exchange-correlation (XC) functional, and it is
{\it a posteriori} corrected for charge inhomogeneity
effects,
using the LDA charge density to
evaluate the exchange-correlation energy in the
generalized gradient approximation (GGA) by Perdew
{\it et al.}\cite{PW91}
The GGA has proven to be of importance for the
calculation of activation energy barriers\cite{Fan}
and barriers of adsorption.\cite{Hammer1,Hammer2}
The dissociation barrier for H$_2$ on Cu(111),
for example, comes out large (order of 0.7 eV)
in the GGA calculation, but is almost zero
in LDA, and only the GGA result conforms with
experiment.\cite{Hammer2,GrossHammer}
We have generated the pseudopotential for Si with
Hamann's\cite{Hamann} scheme, while we use the full
$1/r$ potential for the hydrogen atoms.
The main contribution of the GGA correction to the total
energy is expected to stem from the hydrogen, thus
we do not expect that the usage of an LDA pseudopotential
for Si seriously affects our results.
Calculating the energy gain due to buckling
for the clean Si surface, both the
LDA calculation and our {\it a posteriori} GGA procedure
(in combination with the Hamann LDA pseudopotential)
yield the same result within some meV,
as opposed to GGA corrections of a few hundred meV
for, e.g., the height of the energy barrier of dissociative H$_2$
adsorption.  The {\bf k}-integration\cite{Monkhorst} is performed by using
one special ${\bf k}$-point for the Brillouin zone of the (2$\times$2) cell.
This {\bf k}-point restriction
induces an error of about 0.1 eV.
The cutoff energy defining the plane-waves basis-set
was chosen to be 30 Ry, leading to an estimated
convergence error of about 0.14 eV for
the potential energy surface. To improve the
accuracy of the barrier height the energy of the transition state geometry
was calculated with 4 ${\bf k}$-points
and an energy cutoff of 40 Ry.
The calculated endothermicity of H$_2$ adsorption is
$2.1$ eV per H$_2$ molecule\cite{remark}
(without corrections for zero-point vibrations),
somewhat smaller than the values of $2.4$ eV,\cite{Nachtigall}
$2.9$ eV,\cite{Wu}, and $2.6$ eV\cite{Jing} from previous
cluster calculations.
A detailed presentation of the high-dimensional potential-energy
surface and convergence tests will be published elsewhere\cite{Pehlke}.

The equilibrium structure of the clean Si(001)
surface\cite{Needs,Dabrowski,Wolkow} (see Fig.\ \ref{structure})
is characterized by rows of buckled dimers, resulting
in a p(2$\times$2) or a c(4$\times$2) surface reconstruction.
The buckling is due to the dehybridization of the four $sp^3$
orbitals into three $sp^2$ plus one  $p$ orbital at the ``down'' atom (i.e.,
the
Si dimer atom closer to the bulk), the energetically higher
$p$ orbital is unoccupied and hence the total energy is lower
than for symmetric dimers.\cite{Chadi}
Furthermore, the ``down'' dimer atom tends to push aside its
nearest neighbors in the second layer.
In the p(2$\times$2) structure the second layer atoms
can relax this stress,
which is not possible in the smaller
(1$\times$2) surface unit cell, thus explaining the
preference for structures with alternating buckling
angle.
When the monohydride surface is formed, every dangling
bond of the dimerized Si(001) surface is saturated with
one H atom. The mechanism leading to buckling does not work
any more in this case, hence the dimer bond becomes
parallel to the surface.
Therefore, comparing the initial and final
geometries for a hydrogen molecule dissociatively
adsorbing on a Si(001) surface, it becomes obvious
that the adsorption as well as also the desorption process
have to be accompanied by rather large movements of the
substrate atoms.

The first order desorption kinetics observed in experiment\cite{Hoefer}
is consistent with a pre-pairing mechanism of the
H atoms on the Si surface dimers,
i.e.,  the H$_2$ molecule is formed from two H atoms
that were bond to the same Si dimer prior to desorption.
Thus we do not have to investigate processes with H atoms coming from different
dimers, and we can restrict our geometries to H atoms moving
in the plane spanned by the (001) surface normal and the Si dimer
bond. We calculate the potential energy of an H$_2$ molecule
(with the molecular axis kept parallel to the surface)
as a function of the hydrogen atom separation and the
hydrogen-surface separation. These cuts through the potential energy
hypersurface are parameterized by the displacement of the
center of mass $Y$ of the H$_2$ molecule along the dimer bond;
$(X,Y)=(0,0)$ refers to the center of the (symmetric) Si surface dimer.

First, we freeze the Si atomic positions to those of the
symmetric (1$\times$2) structure.
The Si dimer bond length for the monohydride surface is only by 0.1 \AA{}
larger than for the clean symmetric (1$\times$2)
surface, and the elastic energy of the clean Si(001) surface
corresponding to this expansion amounts to only about $0.1$ eV.
Freezing the Si coordinates to this geometry corresponds
to the physical picture of a ``sudden
desorption event'', i.e., the H$_2$ molecule is assumed to leave the surface
much faster than the Si atoms can relax to the geometry of the clean surface.
The calculated potential energy surface has a large barrier
of about 0.7 eV for the symmetric desorption path (i.e., the
center of mass of the H$_2$ molecule is always above the
center of the surface dimer, and only moves in the
direction perpendicular to the surface) measured with respect to
a free H$_2$ molecule plus a clean symmetric (1$\times$2) surface.
Asymmetric pathways lead to even larger barriers.
This result is in qualitative agreement with Jing and
Whitten's\cite{Jing} cluster calculations, which yielded a
1.15 eV barrier for a similar geometry.

Next we investigate the adiabatic limit, i.e., all Si
atom positions in the topmost three layers are fully relaxed
for every fixed position of the two H atoms. The adsorption energy
barrier is defined with respect to the total energy
of a free hydrogen molecule plus the clean p(2$\times$2)
surface.
The ``adiabatic'' adsorption energy barrier determined
in this way represents the smallest possible barrier.
Paths corresponding to a partial relaxation of the
surface will show a larger, or at best equal, barrier.

Several cuts through the potential energy surface are
shown in Fig.\ \ref{potential}. The optimum desorption
path is asymmetric, and at the transition geometry (see also
Fig.\ \ref{structure}) the H$_2$ molecule is roughly
above the Si dimer atom closer to the bulk.
Close to this geometry, the buckling angle of the
surface Si dimer below the hydrogen molecule
is reduced by about $5^{\rm o}$ with respect to a
value of $\approx 19^{\rm o}$ for the clean p(2$\times$2)
reconstructed surface, and the Si dimer bond length is
slightly increased by about 0.1 \AA.
The adsorption energy barrier height amounts to $\approx 0.3$ eV.
The energy of zero-point vibration
has not been included, however,
as bonds are in general softened at the transition geometry,
we expect that the energy barrier will even be lowered
when vibrational effects are included.

Having found a low adsorption energy barrier path, we tend
to exclude those models for the adsorption and desorption
dynamics of H$_2$/Si(001) that depend on a large energy barrier.
This does not mean that the corresponding physical mechanisms
are absent, but they are not dominant.
When the hydrogen molecule desorbs from the surface, some
energy will be stored in the Si surface vibrations, but this
energy only amounts to $< 0.3$ eV and not about 1 eV.
In fact we calculated that the elastic energy contained in the
Si substrate at the transition geometry is only $\approx 0.15$ eV.
Our results favor the mechanism suggested by Kolasinski
{\it et al.}\cite{KolasinskiPRL,KolasinskiJCP} on the
basis of their experimental findings:
The sticking coefficient is small because the hydrogen
molecules impinging from the gas phase are fast and their
momentum transfer to the substrate is ineffective due to the
mass mismatch, therefore most of them
experience a substantial barrier corresponding to the
geometric configuration of the Si-substrate atoms at the moment
of the interaction. Though we have not calculated the
adsorption energy barriers for such frozen-in configurations
we expect them to be large, similar to the barrier
we calculated for the frozen symmetric surface.
The timescale for scattering of an H$_2$ molecule
can roughly be estimated by the time it takes
a hydrogen molecule at 300 K on the average to transverse
a length of the order of 2 \AA, i.e. $\approx 0.1$ ps.
The angular oscillation of the Si dimer is expected to be
slow ($\lesssim$~3 THz) compared to the optical or acustical
zone-boundary Si bulk phonons, due to the small restoring
forces from bond bending at the surface.
The timescale on which the dimers flip
between their two extremal positions (and thus go through a
configuration resembling a symmetric dimer as in the
monohydride surface structure) was estimated
to be about 10 ps at 300 K,\cite{Dabrowski} which is two orders of magnitude
larger than the scattering time of the H$_2$ molecule impinging on the surface.
On the other hand, hydrogen atoms adsorbed on the surface
stick on the surface until
an appropriate fluctuation of the atomic positions of
the Si surface atoms allows them to desorb and associate
to H$_2$ following a path close to the optimum one with a
small energy barrier.
Hence, for the system H$_2$/Si(001) the adsorption and
desorption dynamics crucially depend on surface atom relaxation.
Microscopic reversibility is not violated, but time-reversed
trajectories are experimentally practically inaccessible as
the coordinates of both the
molecule and the surface Si atoms have to be considered.

After completion of our work we received a preprint of
Kratzer {\it et al.}\cite{Kratzer}. Their density-functional
calculations are very similar to ours and yield a desorption
pathway in close agreement to that discussed above. Their calculated
energy of the transition geometry is slightly higher ($0.5$ eV instead of
$0.3$ eV), which might be due to their smaller super cell.
They assume a (1$\times$2) periodicity of the surface.

We thank K. Kolasinski, E. Hasselbrink, and A. Gro{\ss}{}
for stimulating discussions.



\begin{figure}
\caption{
Atomic structure of, from top to bottom, the clean,
p(2$\times$2)-reconstructed Si(001) surface, the transition state
geometry with a hydrogen molecule close to the Si(001) surface,
and the monohydride surface.}
\label{structure}
\end{figure}

\begin{figure}
\caption{
Potential energy surface of a hydrogen molecule in front of a
Si(001) surface. The contour spacing is 0.2 eV.
The Si substrate has been relaxed for each position of the
hydrogen atoms. The hydrogen molecule is parallel to the surface
and within the plane spanned by the Si surface dimer bond
and the surface normal. The cuts plotted here are parameterized
by the displacement $Y$ of the center of mass of the molecule
along the dimer bond in units of the Si bulk bond length $d_{Si}$.
$Y/d_{Si}=0.5$ corresponds to a
hydrogen molecule coming down onto the surface with the center
of mass roughly above the ``down'' Si-dimer atom (see middle panel
of Fig. 1).}
\label{potential}
\end{figure}


\begin{references}
\bibitem{remark} The endothermicity is defined as the energy difference
        per H$_2$ molecule between the respective ground state configurations,
        i.e., the monohydride surface on the one hand
        and a clean p(2$\times$2)-dimerized
        Si(001) surface plus free H$_2$ molecules on the other hand.
        The endothermicity measured
        with respect to the (1$\times$2) surface
        with symmetric dimers (as used in cluster calculations) would
        amount to $2.3$ eV, closer to the results from cluster calculations.
\end{references}
\end{document}